\documentclass[twocolumn,trackchanges]{aastex701}

\usepackage{xcolor}

\usepackage{booktabs}

\begin{document}

\title{Using CFHT's SITELLE to Probe the Long-Sought Supernova Remnant Shell in the Crab Nebula}

\author[0000-0002-5042-443X]{Lucas V. da Conceição}
\affiliation{Department of Physics and Astronomy, The University of Manitoba, Winnipeg, MB~R3T~2N2, Canada}
\email[show]{daconcel@myumanitoba.ca}  

\author[0000-0001-7068-9702]{Janette Suherli}
\affiliation{Department of Physics and Astronomy, The University of Manitoba, Winnipeg, MB~R3T~2N2, Canada}
\email{suherlij@myumanitoba.ca}

\author[0000-0001-6189-7665]{Samar Safi-Harb}
\affiliation{Department of Physics and Astronomy, The University of Manitoba, Winnipeg, MB~R3T~2N2, Canada}
\email{Samar.Safi-Harb@umanitoba.ca}

\author[0000-0003-2001-1076]{Carter Rhea}
\affiliation{Département de Physique, Université de Montréal, Succ. Centre-Ville, Montréal, Québec, H3C 3J7, Canada}
\email{}

\author[orcid=0000-0002-5044-2988]{Ivo R. Seitenzahl} 
\affiliation{Research School of Astronomy and Astrophysics, Australian National University, Canberra, ACT 2611, Australia}
\email{ivo.seitenzahl@anu.edu.au}

\author[orcid=0000-0002-4794-6835]{Ashley J. Ruiter}
\affiliation{Mathematical Sciences Institute, The Australian National University, Acton, ACT 2601, Australia}
\email{ashley.ruiter@anu.edu.au}

\begin{abstract}

We present deep, wide-field integral field spectroscopy of the Crab nebula obtained with the imaging Fourier transform spectrometer SITELLE at the Canada--France--Hawaii Telescope (CFHT), to search for the long-sought forward shock. Our observations target the coronal line [\ion{Fe}{14}]$\lambda$5303, a tracer of shock-heated gas, over two $11^{\prime} \times 11^{\prime}$ fields that probe projected radii of $\sim$2.4--10~pc west of the pulsar, encompassing the range in which a supernova shell is expected. After data processing and a search over the full field of view, we found no statistically significant  [\ion{Fe}{14}] emission in the surveyed regions. We derive a conservative average surface-brightness upper limit of $\lesssim 3.79 \times 10^{-17}~\mathrm{erg}~\mathrm{cm}^{-2}~\mathrm{s}^{-1}~\mathrm{arcsec}^{-2}$ over three representative annuli. This represents the deepest large-area optical constraint on coronal iron emission beyond the visible nebula from the Crab's putative supernova remnant (SNR) shell. Our results are consistent with scenarios in which the forward shock is expanding into a very low-density medium, the shocked gas is weak or underionized, or the shell lies outside the observed region. This work demonstrates the power of wide-field optical integral field spectroscopy for constraining faint, large-scale structures in SNRs. 

\end{abstract}



\section{Introduction}
\label{sec:intro}

Pulsar wind nebulae (PWNe), also referred to in the literature as filled-center supernova remnants (SNRs) or ``plerions''
\citep[see e.g.,][for a review]{safi2012plerionic, 2006ARA&A..44...17G}, form a distinct subclass of SNRs, shaped by the interaction between a pulsar's relativistic wind and the surrounding medium. The Crab nebula stands as the archetype of this class, and it has been extensively studied across multiple wavelengths \citep{2008ARA&A..46..127H}. Its complex structure includes a bright inner synchrotron nebula visible in X-rays, infrared, and radio, along with an intricate network of optical filaments \citep[e.g.,][]{weisskopf2000,bietenholz2004,martin2021}.

For this class of SNRs, canonical evolution models \citep[e.g.,][]{chevalier1977canonical,truelove1999evolution} predict the presence of a forward shock shell expanding into the interstellar medium (ISM). However, despite decades-long studies, the absence of a supernova shell in $\gtrsim$10-15\% of the Galactic SNR population\footnote{snrcat.physics.umanitoba.ca/SNRtable.php} remains a long-standing gap in our understanding of this class of objects \citep{2012AdSpR..49.1313F, safi2012plerionic}. In particular, out of the 383 known Galactic SNRs, approximately 100 host plerions or plerionic candidates, and nearly half of them ($\sim$52, including the Crab nebula) lack clearly identified outer shells. The absence of detectable forward-shock emission in these systems remains an open problem in the study of PWNe and their evolution. Possible explanations include intrinsically low-energy supernova explosions, expansion into low-density cavities or progenitor wind-blown bubbles, and observational limitations associated with faint shells.

The Crab nebula is widely regarded as the remnant of a core-collapse supernova explosion. However, it also appears anomalous when considered as a young SNR. In particular, the ejecta mass is unusually low  \citep[$4.6 \pm 1.8$~ M$_{\odot}$;][]{fesen1997} and the maximum ejecta velocity is also relatively small \citep[see for instance,][and references therein]{martin2021}, leading to the explosion kinetic energy estimate to be $\sim$10$^{49}-10^{50}$ erg, well below the canonical 10$^{51}$ erg expected for core-collapse supernovae \citep{davidson1985recent}. Together, these properties are atypical of a young core-collapse SNR.

The ``missing'' kinetic energy and mass have been attributed to a yet-undetected outer shell of freely expanding ejecta \citep[e.g.,][]{frail1995does,chevalier1977canonical}. The forward shock is expected to be located at several times the radius of the Crab's visible ejecta \citep{2008ARA&A..46..127H}, corresponding to an approximate radius of $\sim$5~pc from the Crab pulsar. To date, extensive observational searches across multiple spectral bands have been conducted to detect such a shell on these large scales, but no conclusive evidence has been found so far \citep[e.g.,][]{frail1995does,fesen1997,seward2006,2018PASJ...70...14H}.

The forward shock produced by the SN~1054 explosion is expected to interact with the surrounding medium, generating a shell of hot, X-ray emitting gas (e.g., \citealt{chevalier1977canonical,vink2012}). However, to date, there is no such detection. Early ROSAT studies searched for the long-predicted SNR shell of the Crab nebula but found no clear evidence of extended emission \citep{mauche1989x, predehl1995x}. Using deep \textit{Chandra} observations, \cite{seward2006} searched for thermal X-ray emission from within and around the nebula and found no significant thermal X-ray emission from an outer shell, arguing against a clearly detected supernova shell in X-rays. More recently, the Hitomi collaboration searched for thermal X-ray emission or absorption lines from an unseen shell and similarly found no thermal shell signature (\citealt{2018PASJ...70...14H}). Combining the Hitomi data with earlier X-ray upper limits, the X-ray emitting plasma was constrained to about $\lesssim$ 1 $M_{\odot}$.

In the optical band, observations have identified a diffuse optical halo extending up to 14~arcmin from the pulsar \citep{murdin1981} and a sharp [\ion{O}{3}] boundary \citep{gull1982}, later interpreted as the Rayleigh-Taylor instability interface between the synchrotron nebula and the freely expanding ejecta, commonly referred to as the [\ion{O}{3}] ``skin'' \citep[e.g.,][]{sankrit1997}. Additionally, high-velocity optical line emission has been detected in the central region, with velocities up to 3600~km\,s$^{-1}$ \citep{clark1983three,lawrence1995three}. \citet{clark1983three}, in particular, reported high-velocity [\ion{O}{3}] emission extending to radial velocities of approximately $+3600$~km~s$^{-1}$ for the $\lambda5007$ component and $-2400$~km~s$^{-1}$ for the $\lambda4959$ component.

A broad, blueshifted \ion{C}{4}~$\lambda1550$ absorption line observed toward the pulsar, reaching a maximum velocity of approximately 2500~km~s$^{-1}$ \citep{sollerman2000}, provides the only hint to date for an outer shell enclosing the Crab nebula. Furthermore, studies of [\ion{O}{3}]$\lambda\lambda4959,5007$ and \ion{Ca}{2} $\lambda\lambda3934,3968$ along the line of sight have constrained the primary shell's velocity range between $-1125$ and $1075$ $(\pm 125)$~km\,s$^{-1}$ \citep{lundqvist2012}. Despite these findings, direct observational confirmation of the Crab's missing shell remains elusive. 

While some studies initially reported the presence of a $6' \times 14'$ halo in deep H\,$\alpha$ imaging \citep{murdin1981} and a faint, narrow H\,$\beta$ emission beyond the northern edge of the remnant \citep{murdin1994stellar}, which could explain the missing kinetic energy, \citet{fesen1997} found no evidence of an optical halo in H\,$\alpha$ and concluded that the faint H\,$\beta$ emission is likely diffuse Galactic emission.  
Similarly, radio observations have also failed to provide evidence for the existence of an extended outer shell. Early studies searching for large-scale radio features associated with the Crab nebula did not reveal any significant emission beyond the well-known synchrotron nebula \citep{velusamy1984radio,velusamy1992multifrequency,frail1995does}. 

While forward shocks in SNRs are primarily studied at X-rays, and to a lesser extent in radio and infrared, their signature can also be traced in the optical through high-ionization coronal Fe lines such as [\ion{Fe}{14}]$\lambda 5303$. \citet{shklovskii1967} first proposed that extremely hot plasma in SNRs should produce such coronal emission. These lines trace the forward-shocked gas and, being observable from the ground, provide spatially resolved diagnostics at resolutions far exceeding those typically achievable in X-ray observations. Moreover, recent observations with the \textit{James Webb Space Telescope} (JWST) have provided new insights into the nucleosynthetic products of the Crab Nebula. Spatially resolved measurements of the Ni/Fe abundance ratio have revealed significant chemical inhomogeneities and provided new constraints on the explosion mechanism and progenitor evolution. These results demonstrate that the nebula remains an important laboratory for investigating both the dynamics and chemical evolution of core-collapse supernova remnants, motivating continued searches for observational signatures of its long-sought forward shock.

Optical coronal iron emission has been detected in a number of SNRs. In our Galaxy, it has been reported in remnants such as IC~443 \citep{woodgate1979,Sauvageot1990} and Puppis~A \citep{1979Natur.281..656I,lucke1979forbidden}. Similar emission has also been observed in the Magellanic Clouds remnants, including the core-collapse SNRs N49 \citep{Murdin1978,Dopita2016_fe}, N132D \citep{Dopita2018}, 1E~0102.2--7219 \citep{vogt2017fe}, and 0540--69.3 \citep{Tenhu2025}, as well as the Type~Ia remnants SNR~0519--69.0, SNR~0509--67.5, and N103B \citep{Seitenzahl2019}. These results establish coronal iron lines as a robust diagnostic of shock-heated gas and justify their use in the search for the Crab’s elusive forward shock.

The presence of coronal [\ion{Fe}{14}]$\lambda$5303 emission in the Crab Nebula has been discussed as a potential tracer of hot gas associated with the outer shock. 
\cite{lucke1979forbidden} examined the soft X-ray bright regions in the northeast and south of the synchrotron emission in the Crab nebula, where the inferred plasma temperatures reach 5 $\times$ 10$^6$~K. At such temperatures, most iron is expected to be ionized beyond the level that can produce strong optical forbidden lines. 
Nevertheless, they considered the possibility that cooler gas might produce detectable [\ion{Fe}{14}] emission and derived 
a flux upper limit of $8\times10^{-14}$~erg~cm$^{-2}$~s$^{-1}$ 
within a 3~arcmin$^2$ aperture. This corresponds to an average surface brightness of $7.41\times10^{-18}$~erg~cm$^{-2}$~s$^{-1}$~arcsec$^{-2}$.
However, this emission is expected to be highly structured and spatially localized in thin filamentary regions with a small volume filling factor, rather than uniformly distributed across the nebula 
\citep[e.g.][]{vogt2017fe,lundqvist2012}. 
The surface brightness of these localized, limb-brightened regions could therefore be significantly higher than the implied average value, as spatial averaging makes the analysis most sensitive to emission that is extended over a substantial fraction of an extraction region. In addition to the annular spectral extraction, the continuum-subtracted cubes and [\ion{Fe}{14}] intensity maps were examined across the full observed field for coherent localized, filamentary, or shell-like emission. No convincing structure was identified. However, because this full-field examination was not converted into a completeness-calibrated blind-search limit for arbitrary source sizes and morphologies, we do not assign a general surface-brightness upper limit to localized emission.

In this work, we use SITELLE\footnote{https://www.cfht.hawaii.edu/Instruments/Sitelle/}, the optical imaging Fourier transform spectrometer mounted on the 3.6-m Canada--France--Hawaii Telescope (CFHT). Combining a wide field of view with imaging spectroscopy at adjustable spectral resolution across the optical domain \citep{drissen2019sitelle}, SITELLE is well suited to probing the faint emission surrounding the visible nebula and searching for signatures of the predicted forward shock.

\section{Observations, data reduction,
and post-processing} 
\label{sec:sec2}

\begin{figure}[h]
\plotone{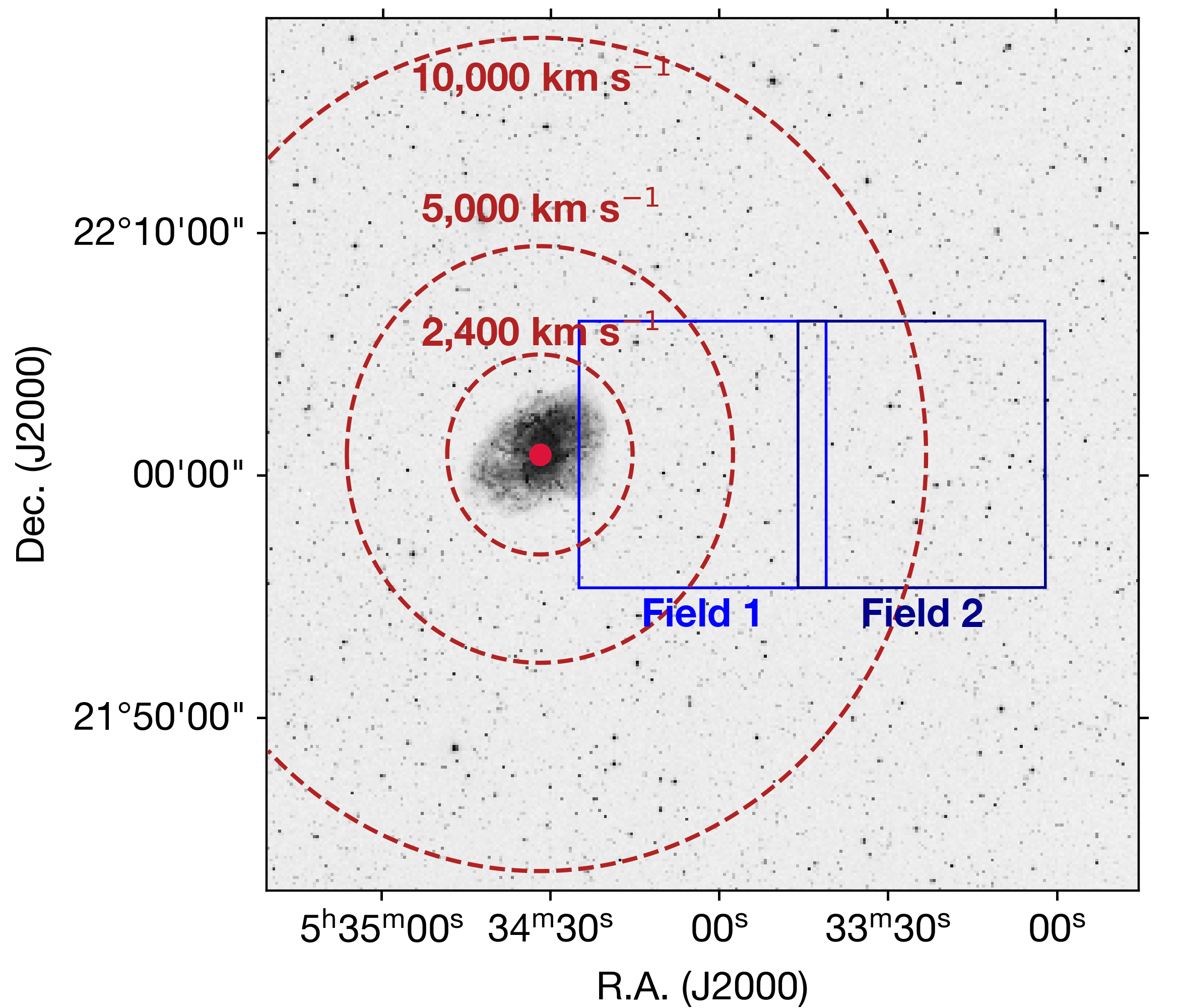}
\caption{Sky image of the Crab nebula (from the Digitized Sky Survey 2) showing the location of our SITELLE observations. The blue squares, each covering 
$11\arcmin \times 11\arcmin$, indicate the footprints of Field~1 and Field~2, with a small region of overlap between them. The red dashed concentric circles mark projected radii of 2.4, 5, and 10~pc from the pulsar (corresponding to $\sim$4–17~arcmin at a distance of 2~kpc), illustrating the range of plausible locations of the forward shock. These radii are motivated by simple free-expansion scaling for velocities of $2400$, $5000$, and $10{,}000$~km~s$^{-1}$ over the $\sim$1 kyr age of the remnant. The red dot marks the position of the Crab pulsar.
}
\label{fig:fig1}
\end{figure}

Observations of the Crab nebula were conducted on the nights of October 24, 2023, and January 1, 2024, as part of program 23BC06 (P.I.: J. Suherli). The datasets consist of two datacubes with a total exposure time of 5.71~hours each (including overheads) obtained with the C3 filter (511--556~nm), using 53 interferometer steps with an exposure time of 392.3~seconds per step. The cubes account for observations of two fields of $11^{\prime} \times 11^{\prime}$ (see Figure~\ref{fig:fig1}), centered at 
$05^{h}34^{m}02^{s}.90;+22^{\circ}00^{\prime}52.1^{\prime\prime}$ and $05^{h}33^{m}23^{s}.9;+22^{\circ}00^{\prime}52.1^{\prime\prime}$ [J2000]. These fields were selected to target the [\ion{Fe}{14}]$\lambda5303$ coronal line and to probe the region where the Crab's missing shell is expected, at projected radii of $\sim$2--10~pc from the pulsar. This range corresponds to $\sim$4--17~arcmin on the plane of the sky at an assumed distance of 2~kpc. These radii are motivated by simple free-expansion scaling for ejecta velocities in the range $\sim$2400--10000~km~s$^{-1}$ over the $\sim$1000 year age of the remnant, and should be regarded as order-of-magnitude estimates rather than precise dynamical predictions.

The data were processed using the ORBS pipeline \citep{martin2015orbs}, which performs standard de-trending, cube reconstruction, and corrections for image alignment and instrumental effects. Wavelength calibration was obtained using a HeNe laser cube, and spectrophotometric calibration was based on standard star observations.
A detailed description of the data reduction process can be found in \citet{martin2016optimal}.

To account for Galactic extinction along the line of sight, both data cubes were corrected using the \textsc{Python} package \texttt{extinction}, which applies the \citet{fitzpatrick1999} correction curve with $R_{V} = 3.1$, adopting a color excess of $E(B-V) = 0.52$ \citep{sollerman2000}. Local sky subtraction was performed using regions within each cube that are relatively free of emission across 
the full spectral range, allowing for the removal of background and sky contributions from the extracted spectra.

\begin{figure*}[ht]
\plotone{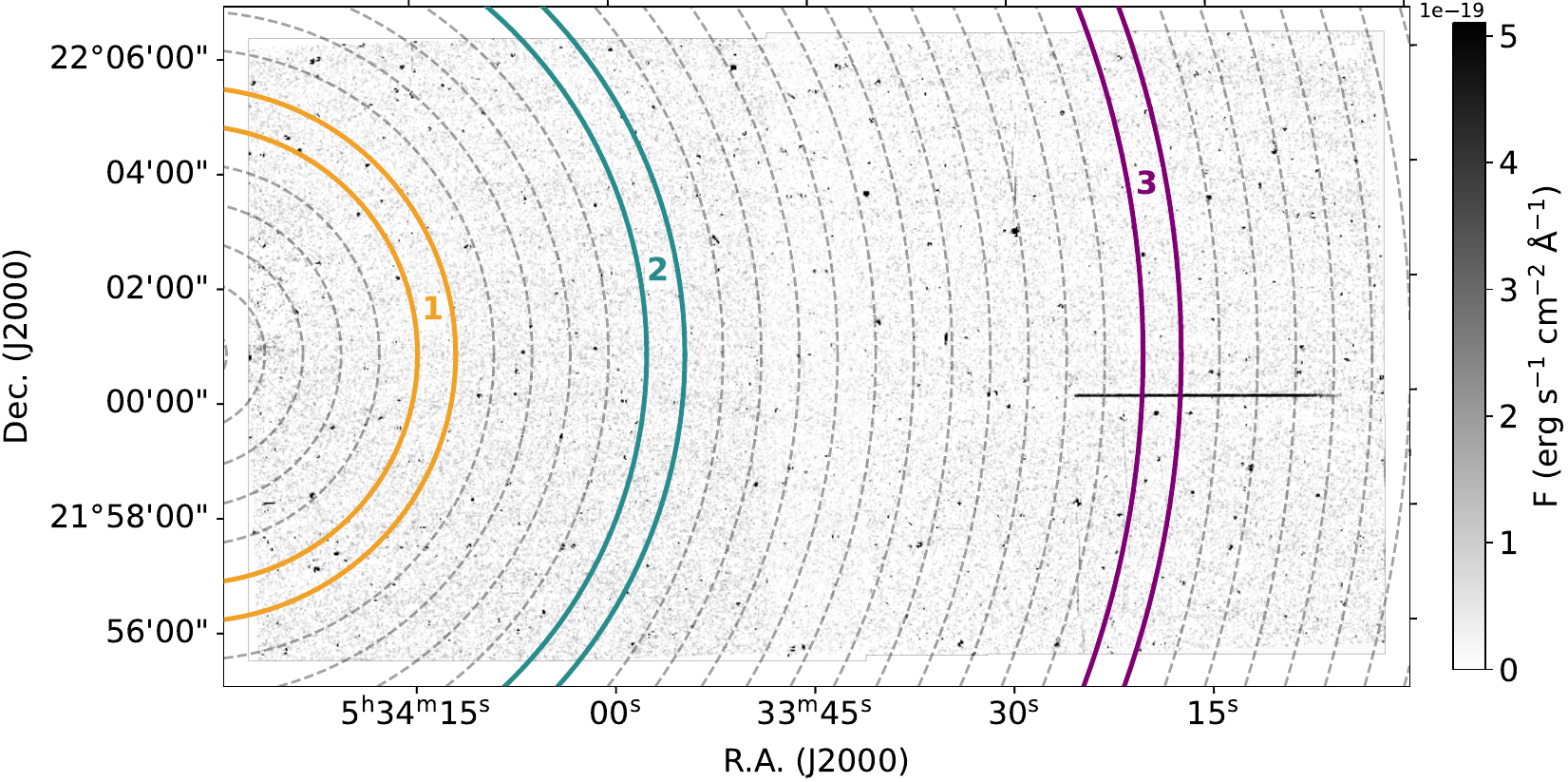}
\caption{Intensity map of [\ion{Fe}{14}]$\lambda5303$ coronal line from the SITELLE observations of Field~1 and Field~2 combined, 
integrated from $-500~\mathrm{km\,s^{-1}}$ to $+500~\mathrm{km\,s^{-1}}$. The image is shown in greyscale with a common intensity scale. The colored curves indicate the projected locations of three annuli centered on the Crab's pulsar at radii of 2.4 and 5~pc in Field~1 (orange [1] and teal [2], respectively), and 10~pc in Field~2 (magenta [3]), each with a width of 40~arcsec. No obvious filamentary or shell-like emission associated with [\ion{Fe}{14}] is detected in either field. Note that the horizontal line that appears on Field 2 is an instrumental artifact. The Crab pulsar and nebula are on the left edge of the figure.
}
\label{fig:fig2}
\end{figure*}

The transmission of the SITELLE C3 filter decreases rapidly near the edges of the bandpasses, resulting in a drastic decline in throughput on both ends \citep{Martin2021_dataRed}. To avoid introducing artifacts and uncertainties, the spectral cube was trimmed where the transmission drops significantly. 
To enhance the visibility of faint nebular emission features against the underlying continuum emission, we performed continuum subtraction on both data cubes. The continuum in each spatial pixel (spaxel) was modeled using 
the Locally Weighted Scatterplot Smoothing algorithm \citep[LOWESS;][]{cleveland1979}, 
which provides a non-parametric and locally adaptive fit that is robust to outliers 
through an iterative process of local regression. The resulting continuum model was then subtracted from the original spectra. A similar implementation for MUSE (the integral field spectrograph mounted on the Very Large Telescope) data was detailed in \citet{vogt2017fe}.

\section{Results} 
\label{sec:sec3}

\subsection{Search for the [\ion{Fe}{14}] Emission Line}


Figure~\ref{fig:fig2} shows the intensity map of [\ion{Fe}{14}]$\lambda$5303 constructed from the SITELLE data cubes of both Field~1 
and Field~2 
combined, 
integrated 
over $-500 \le v_{\text{los}} \le +500~\mathrm{km\,s^{-1}}$. 
We divided the observed field into concentric annuli 
centered on the Crab pulsar to search for any [\ion{Fe}{14}] emission. Each annulus has a 40~arcsec width, and the colored annuli in Figure~\ref{fig:fig2} indicate the projected locations at radii of 2.4~pc (yellow), 5~pc (teal), and 10~pc (magenta). 

No statistically significant [\ion{Fe}{14}]$\lambda$5303 emission is detected across the observed SITELLE field. Neither the integrated intensity maps nor the extracted spectra from the concentric annuli reveal coherent filamentary, shell-like, or localized structures associated with coronal iron emission. 
The absence of detectable [\ion{Fe}{14}] emission indicates that gas containing a sufficient fraction of Fe$^{13+}$ is either absent or produces emission below the sensitivity limits of the present observations. Possible physical explanations for this non-detection are discussed in Section~\ref{sec:discussion}.

\subsection{Flux Upper Limits}

\begin{figure*}
\plotone{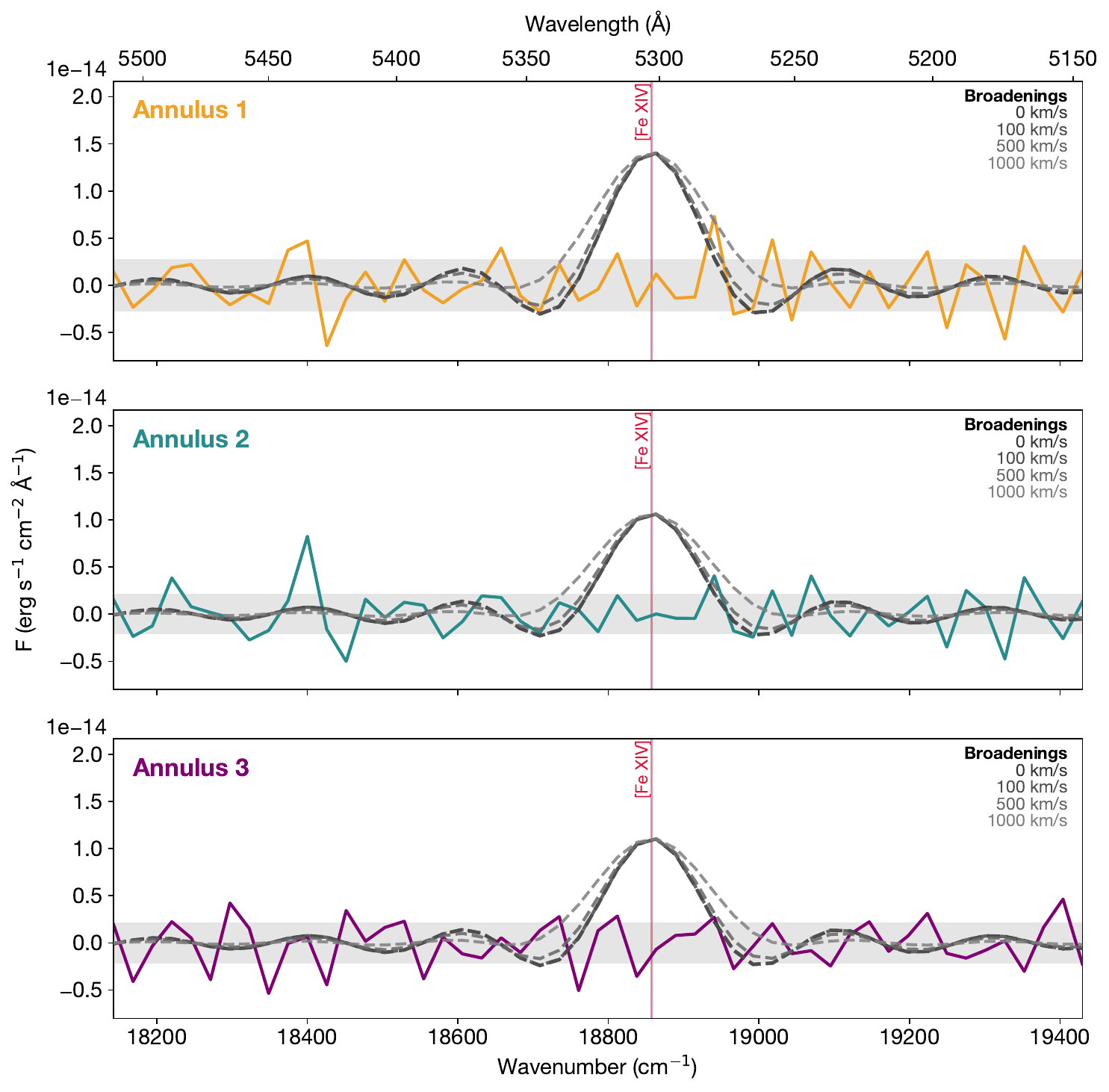}
\caption{ Continuum-subtracted integrated spectra (colored curves) extracted from the three annular regions highlighted in Figure~\ref{fig:fig2} (1, 2, and 3). The dashed vertical red line marks the rest wavelength 
of [\ion{Fe}{14}]$\lambda5303$. The grey curves show 
model sincgauss line profiles corresponding to different assumed intrinsic velocity dispersions ($\sigma_v = 0$--1000~km~s$^{-1}$), with 
their peak amplitudes scaled to the adopted $5\sigma$ detection threshold, where $\sigma$ is the rms noise measured from nearby line-free continuum regions of the continuum-subtracted spectrum. The shaded grey area indicates the $\pm1\sigma$ rms noise level used to define the $5\sigma$ detection threshold.
}
\label{fig:fig3}
\end{figure*}

For an ideal FTS, the instrumental line shape (ILS) is described by a sinc function. The width of the sinc profile, $\Delta w$ (in cm$^{-1}$), is determined by the maximum optical path difference (MPD), such that $\Delta w=1/2\text{MPD}$. The full width at half maximum (FWHM) of the sinc ILS is related to $\Delta w$ through $\text{FWHM} = 1.20671 \Delta w$, which in turn defines the spectral resolution through
\begin{equation}
    R = \frac{\text{wvn}}{\text{FWHM}}, 
\end{equation}
where wvn denotes the wavenumber in cm$^{-1}$ \citep{Kauppinen2001}. To avoid confusion with the standard deviation $\sigma$ used for the noise estimate in this work, we do not adopt the wavenumber notation used by \citet{martin2016optimal}.

Astrophysical emission lines are intrinsically broadened by physical processes such as thermal and turbulent motions that are commonly described by Gaussian profiles. The observed SITELLE line profile therefore can be modeled as the convolution of the instrumental sinc ILS and a Gaussian broadening profile, producing a sincgauss profile \citep{martin2016optimal}. In this work, the sincgauss profiles were evaluated using the numerically stable Dawson-function formulation presented by \citealt{martin2016optimal} (see Equation~(15) therein).

To estimate the flux upper limits for the non-detection of [\ion{Fe}{14}] emission, we modeled the expected line profile at the rest wavelength of [\ion{Fe}{14}]$\lambda$5303. The instrumental sinc width $\Delta w$ was fixed by the SITELLE spectral resolution of our observations, and we also considered intrinsic Gaussian velocity broadenings of [0, 100, 500, 1000]~km\,s$^{-1}$. 
A detection was defined as a peak amplitude exceeding $5\sigma$, where $\sigma$ is the standard deviation (rms noise) measured from nearby line-free regions of the continuum-subtracted integrated spectrum for each annulus. The synthetic sincgauss profiles were scaled to this detection threshold and subsequently integrated over wavenumber space to derive the corresponding flux upper limits for each assumed intrinsic velocity broadening.

Figure~\ref{fig:fig3} compares the expected sincgauss line profiles with the integrated spectra extracted from projected radii of approximately 2.4, 5, and 10~pc from the Crab pulsar. The three annuli were selected to sample physically motivated inner, intermediate, and outer locations for the putative forward shock. Under the simplified assumption of free expansion over the Crab’s $\sim$ 1000 yr age, the innermost radius of 2.4 pc corresponds to a mean expansion velocity of approximately 2400 km~s$^{-1}$. This value is comparable to the highest blueshifted optical ejecta velocity reported for the Crab from the [O III]$\lambda$4959 line emission reaching approximately $-$2400 km~s$^{-1}$ in \citet{clark1983three}. The intermediate radius of 5 pc corresponds to a mean free-expansion velocity of approximately 5000 km~s$^{-1}$, which is representative of forward-shock velocities measured in young SNRs. Finally, the 10 pc annulus corresponds to a substantially higher mean expansion velocity of approximately 10000 km~s$^{-1}$ and was included as an outer bracketing case, sampling a rapidly expanding, weakly decelerated shell near the outer boundary of the SITELLE spatial coverage. Thus, the three annuli provide representative inner, intermediate, and outer sampling scales rather than an exhaustive set of possible shell locations. The dashed vertical red line marks the rest wavelength of [\ion{Fe}{14}], while the grey curves illustrate the expected sincgauss profiles for different assumed intrinsic velocity broadenings. Given the spectral resolution and SITELLE setup of these observations, profiles with intrinsic velocity broadenings of $\lesssim 500$~km~s$^{-1}$ are indistinguishable from the unresolved instrumental sinc profile.

The integrated flux upper limits and corresponding average surface-brightness limits are summarized in Table~\ref{tab:fe14_ul}. The quoted values correspond to the narrow-line case assuming only instrumental broadening. These surface-brightness limits were computed using the effective extraction area within the combined field footprint. Among the three annuli, the most stringent average surface-brightness limit is S$_{\text{UL}} \lesssim 3.79 \times 10^{-17}~\mathrm{erg}~\mathrm{cm}^{-2}~\mathrm{s}^{-1}~\mathrm{arcsec}^{-2}$. The limits in Table~\ref{tab:fe14_ul} should be interpreted as representative average surface-brightness constraints at three physically motivated projected radii within the observed western fields. They do not constitute an azimuthally complete constraint on a circular shell, because only the portions of the annuli intersecting the SITELLE footprints were observed. Moreover, the annular extraction is optimized for approximately shell-like emission.

\begin{table*}
\centering
\caption{
Flux upper limits (F$_{\text{UL}}$) for [\ion{Fe}{14}]$\lambda$5303 emission derived from the integrated spectra extracted from the annulus corresponding to the projected radii of approximately 2.4, 5, and 10~pc from the Crab pulsar. The values correspond to the narrow line case, assuming only instrumental broadening. Surface brightness limits (S$_{\text{UL}}$) were computed using the effective extraction area within the combined field footprint.
}
\begin{tabular}{lcccc}
\toprule
\addlinespace[1.5pt]
\toprule
~Region & Radius & Area & F$_{\text{UL}}$ & S$_{\text{UL}}$ \\
 & (pc) & (arcsec$^2$) & 
 (erg\,s$^{-1}$\,cm$^{-2}$) & (erg\,s$^{-1}$\,cm$^{-2}$\,arcsec$^{-2}$) \\
\midrule
~1  & 2.4  & 27,614.8 & 1.43 $\times$ 10$^{-12}$ & 5.17 $\times$ 10$^{-17}$ \\
~2 & 5.0  & 28,456.9 & 1.08 $\times$ 10$^{-12}$ & 3.79 $\times$ 10$^{-17}$ \\
~3 & 10.0 & 26,594.7 & 1.12 $\times$ 10$^{-12}$ & 4.22 $\times$ 10$^{-17}$ \\
\bottomrule
\end{tabular}
\label{tab:fe14_ul}
\end{table*}

\section{Discussion and Conclusions}
\label{sec:discussion}

Our SITELLE observations, 
targeting the coronal [\ion{Fe}{14}]$\lambda$5303 emission line in the western region beyond the visible nebula,
yield no statistically significant detection of a forward shock over the searched radial range (2.4--10~pc). 

The absence of detectable [\ion{Fe}{14}] emission suggests that either coronal-temperature ionized gas is not present in sufficient density, or that the forward shock in the Crab nebula is too weak to collisionally excite the [\ion{Fe}{14}] line above our detection threshold. 

Our spatially averaged surface-brightness upper limit is the deepest large-area optical constraint to date for the [\ion{Fe}{14}] line beyond the visible nebula in the surveyed regions.
Possible explanations for this non-detection include low iron emissivity 
in the shocked medium, a low-density environment, or a forward shock located outside the SITELLE field. We discuss these possibilities below, and then place our results in the broader context of a low-energy supernova explosion.

  
\textbf{Non-equilibrium ionization time-scale:} 

One possible explanation for the absence of detectable [\ion{Fe}{14}] emission is that the shocked plasma has not yet reached ionization equilibrium. In a collisionless shock, the bulk kinetic energy of the upstream flow is rapidly thermalized at the shock front, initially heating the positive ions primarily. The electrons subsequently gain energy through Coulomb interactions with these ions and progressively ionize the heavy elements toward higher charged states \citep{Seward2010}. Consequently, immediately behind the shock, the ionization state may lag behind the electron temperature, producing a non-equilibrium ionization (NEI) plasma \citep{vink2012}.

The ionization timescale governs the evolution of the ionization state,
\begin{equation}
\tau = n_{e} t,
\end{equation}
where (n$_{e}$) is the electron density and $t$ is the time elapsed since the gas was shocked. Young SNRs are usually not yet in equilibrium and commonly exhibit ionization timescales of approximately
\begin{equation}
n_{e} t \sim 1\times~10^{10}-3\times10^{11}~cm^{-3}s,
\end{equation}
whereas collisional ionization equilibrium is generally approached for
\begin{equation}
n_{e} t \gtrsim 3\times10^{12}~cm^{-3}s
\end{equation}
\citep{vink2012}.

The NEI calculations of \cite{2024ApJ...976..180O}, for instance, indicate that Fe$^{13+}$, the ion responsible for the [\ion{Fe}{14}]~$\lambda5303$ line, reaches its maximum abundance near
\begin{equation}
n_{e} t \sim 10^{10}~cm^{-3}s.
\end{equation}
For a characteristic timescale comparable to the age of the Crab Nebula,
($t \sim 10^{3}~yr \approx 3.2\times10^{10}~s$), this ionization timescale corresponds to an electron density of

\begin{equation}
n_{e} \sim \frac{10^{10}~cm^{-3}s}{3.2\times10^{10}s}\sim 0.3~cm^{-3}.
\end{equation}

Therefore, if the post-shock electron density is substantially below approximately $0.3~cm^{-3}$, the plasma may not yet have developed a significant Fe$^{13+}$ population, even if the shock has heated the gas to temperatures at which this ion could eventually become abundant.

Electron-ion temperature non-equilibrium may provide an additional source of suppression. Immediately behind a collisionless shock, the ion temperature can substantially exceed the electron temperature, and Coulomb equilibration may proceed slowly in low-density plasma. Because both the collisional ionization of iron and the collisional excitation of the [\ion{Fe}{14}]~$\lambda5303$ transition are driven by electrons, a comparatively low electron temperature can delay the production of Fe$^{13+}$ and reduce the excitation rate of the line. Thus, the absence of detectable [\ion{Fe}{14}] emission does not necessarily imply the absence of shock-heated gas. It may instead indicate that the plasma remains underionized, that the electrons have not fully equilibrated with the ions, or that both effects are operating simultaneously.

\textbf{Low-density surroundings:} Another plausible explanation is that the forward shock is propagating into an unusually tenuous ambient medium. In this case, the expected [\ion{Fe}{14}] emission may be suppressed for several independent reasons. First, the line emission scales with the emission measure, $EM=\int n_{e}^{2}dl$, so even shock-heated plasma can produce intrinsically weak emission if the post-shock density is low. Second, as discussed in the previous section, low electron densities imply small ionization timescales ($n_{e}t$), delaying the production of Fe$^{13+}$ and reducing the [\ion{Fe}{14}] emissivity under non-equilibrium ionization conditions. Consistent with this scenario, no forward shock or outer shell has been detected at radio, optical, or X-ray wavelengths \citep{frail1995does, 2008ARA&A..46..127H,seward2006, 2018PASJ...70...14H}, and the nebula exhibits only weak deceleration, remaining close to free expansion \citep{martin2021}. This implies that the swept-up mass is still negligible compared to the ejecta mass, as expected in a low-density surrounding medium. Furthermore, electron-ion temperature equilibration proceeds more slowly in low-density plasmas, potentially maintaining electron temperatures below the ion temperature and thereby reducing the efficiency of collisional excitation.

Differently than shocks propagating through dense media, a forward shock expanding into a tenuous environment is expected to remain largely non-radiative. Because the radiative cooling time increases approximately inversely with density (see, for instance, \citealt{vink2012}), thermal energy is lost only slowly, allowing the shocked gas to maintain a high post-shock temperature over timescales much longer than the age of the Crab Nebula. Consequently, the absence of detectable [\ion{Fe}{14}] emission should not be interpreted as evidence that the forward shock has significantly cooled or weakened. Instead, the shock may remain both fast and hot while producing little observable coronal emission owing to the combined effects of low emission measure, delayed ionization, and incomplete electron-ion thermal equilibration.


Moreover, the absence of strong shock-interaction signatures 
would be consistent with the Crab nebula evolving within a tenuous environment, possibly shaped by a pre-supernova wind cavity. 
This scenario may also be plausible given the Crab's location $\sim$200 pc (Galactic latitude of $-5.7844$) below the Galactic plane, where the hydrogen column density is expected to be lower overall \citep[see][and references therein]{kalberla2009}.
Such a scenario would also be consistent with the absence of X-ray emission beyond the pulsar wind nebula.

\textbf{Shell beyond the field-of-view:}
 
The SITELLE field of view covers only a limited portion of the environment surrounding the Crab nebula on the western side, and therefore cannot exclude the presence of extended, low-surface-brightness emission at other azimuthal directions (see Figure~\ref{fig:fig1}), or possibly at larger distances beyond 10~pc. If such a shell exists, its emission may lie entirely outside the observed fields, as observed in the [\ion{Fe}{14}]$\lambda5303$ coronal iron line emission from the shocked gas in the Crab-like SNR 0540--69.3 \citep{Tenhu2025}, where the emission is predominantly detected toward the southwestern side of the nebula, with only faint or no detectable emission elsewhere. Future wide-field and sensitive mosaic observations covering the full area around the Crab will be essential to robustly test this scenario and to search for faint coronal-line emission associated with a putative forward shock.

Our derived upper limit 
on the [\ion{Fe}{14}]$\lambda 5303$ surface brightness of 
3.79~$~\times~10^{-17}$~erg~cm$^{-2}$~s$^{-1}$~arcsec$^{-2}$
primarily constrains the amount of shocked gas at coronal temperatures, rather than directly the explosion energy itself. The expected [\ion{Fe}{14}] luminosity depends on the shock velocity, ambient density, ionization state, iron abundance, and emitting volume. Therefore, this non-detection disfavors models predicting a sufficiently dense and hot (several $10^6$~K) shell bright in coronal iron emission, but it does not uniquely determine the kinetic energy of the supernova explosion.  Within this broader context, a low-energy explosion ($\sim$10$^{49}$--10$^{50}$~erg) remains a viable scenario for the lack of a detectable forward shock around the visible nebula. Such a scenario was originally proposed by \citet{1982Natur.299..803N}, advancing the view that the Crab resulted from the collapse of an electron-degenerate 
O-Ne-Mg core of an 8-10 M$_{\odot}$ star (see also \citealt{yang2015,kitaura2006}). The lower energy (1-2 orders of magnitude lower than a canonical supernova explosion) would produce a weaker, fainter forward shock whose optical and X-ray emission remain below current detection limits. However, it is important to emphsize that, although our non-detection is consistent with a low-energy explosion scenario, it does not necessarily constitute evidence in favour of such an interpretation, as the present observations alone cannot distinguish between a low-energy explosion and the several other plausible explanations discussed above. Consequently, our observations primarily constrain models that predict detectable [\ion{Fe}{14}] emission within the surveyed region, rather than uniquely constrain the explosion energetics.

In summary, we do not detect statistically significant [\ion{Fe}{14}]~$\lambda 5303$ emission in the western region probed beyond the visible Crab nebula. 
We have placed deep, large-area surface-brightness constraints on coronal emission from a putative forward shock within projected radii of $\sim$~2.4--10~pc west of the Crab pulsar.
This non-detection is consistent with scenarios invoking weak or underionized shocked gas, low iron emissivity, and expansion into a low-density medium. Future progress will require deeper, wider-field IFU observations targeting multiple shock-sensitive lines (e.g., [\ion{Fe}{10}], [\ion{Fe}{11}], [\ion{O}{6}]), combined with more sensitive and wide radio, optical, and X-ray searches,  with next-generation instruments like the Square Kilometer Array (\textit{SKA}, \citealt{2009IEEEP..97.1482D}), Extremely Large Telescopes (ELTs, see e.g. \citealt{2026arXiv260520305H}), the Advanced X-ray Imaging Satellite (\textit{AXIS}, \citealt{2023SPIE12678E..1ER}), and the New Advanced Telescope for High-ENergy Astrophysics (\textit{NewAthena}, \citealt{2025NatAs...9...36C}). Observations covering other regions around the Crab (see Figure~\ref{fig:fig1}) or extending beyond 10~pc from the pulsar would help rule out the presence of a faint shell beyond the visible nebula. Ultimately, constraining or detecting the Crab's forward shock will shed light not only on this iconic remnant’s formation but also on the diversity of shell-less PWNe.



\begin{acknowledgments}

In this study, the preliminary analysis of the data cubes was performed using LUCI \citep{rhea2019,Rhea2021_updatesPaper}, a general-purpose fitting pipeline built with PYTHON specifically to handle SITELLE IFU data cubes. In addition, we made use of NASA's Astrophysics Data System (ADS) and of SNRcat, the High-Energy Catalog of Supernova Remnants hosted at the University of Manitoba \citep{2012AdSpR..49.1313F}.   

 L.V.C., J.S., and S.S.H. acknowledge support for this research from the Natural Sciences and Engineering Research Council of Canada (NSERC) through the Discovery Grants and the Canada Research Chairs programs, and from the Canadian Space Agency.
We thank the CFHT technical staff, particularly Beno\^{\i}t Epinat, for their support of our observation. We thank the referee for their useful comments that helped improve the clarity and presentation of our work.
\end{acknowledgments}

\facilities{CFHT: SITELLE}

\software{
Astropy \citep{astropy2013,astropy2018}, SciPy \citep{virtanen2020scipy},
NumPy \citep{harris2020numpy},
Matplotlib \citep{hunter2007},
Photutils \citep{photutils2016},
LUCI \citep{rhea2019,Rhea2021_updatesPaper},
APLpy \citep{robitaille2012},
Extinction v0. 3.0 \citep{2016zndo....804967B}, TQDM \citep{da2019tqdm}, Pyregion \citep{pyregion}, Statsmodels \citep{seabold2010statsmodels}.
}




\bibliography{references}{}
\bibliographystyle{aasjournalv7}



\end{document}